# Entropic and Near-Field Improvements of Thermoradiative Cells

Wei-Chun Hsu, Jonathan K. Tong, Bolin Liao, Yi Huang, Svetlana V. Boriskina,* and Gang Chen*

*Department of Mechanical Engineering, Massachusetts Institute of Technology, Cambridge, Massachusetts 02139, USA*

**Abstract**

A p-n junction maintained at above ambient temperature can work as a heat engine, converting some of the supplied heat into electricity and rejecting entropy by interband emission. Such thermoradiative cells have potential to harvest low-grade heat into electricity. By analyzing the entropy content of different spectral components of thermal radiation, we identify an approach to increase the efficiency of thermoradiative cells via spectrally selecting long-wavelength photons for radiative exchange. Furthermore, we predict that the near-field photon extraction by coupling photons generated from interband electronic transition to phonon polariton modes on the surface of a heat sink can increase the conversion efficiency as well as the power generation density, providing more opportunities to efficiently utilize terrestrial emission for clean energy. An ideal InSb thermoradiative cell can achieve a maximum efficiency and power density up to 20.4 % and 327 $Wm^{-2}$, respectively, between a hot source at 500K and a cold sink at 300K. However, sub-bandgap and non-radiative losses will significantly degrade the cell performance.

**Keywords**: Energy conversion, near-field radiation, phonon polariton modes, thermoradiative cell, entropy, mid-infrared emission, spectral selectivity.

*Correspondence and requests for materials should be addressed to either S.V.B. or G.C. (emails: sborisk@mit.edu, gchen2@mit.edu)

**Introduction**

Low-grade heat is omnipresent in industrial processes, vehicular engines, and power electronics.[1-2] Many technologies such as solid-state thermoelectric energy conversion,[3-5] organic Rankine cycles,[1,6] and thermally regenerative electrochemical cycle[7-8] are pursued to harvest energy from low temperature heat sources. Semiconductor p-n junctions, on the other hand, have been used to harvest photons from high temperature heat sources to generate electricity, such as photovoltaic (PV) and thermophotovoltaic devices.[9-15] In these devices, photons from external heat source enter the devices to generate electron-hole pairs and carry in entropy. Entropy of the incoming photons and entropy generated in the energy conversion process will be carried away by photons emitted during radiative recombination and via heat rejected to the environment.

Recently, the potential of using a p-n junction directly as a heat engine, coupling heat into the p-n junction by conductive or convective heat transfer from a heat source and rejecting the entropy via thermal radiation, has been explored for harvesting low temperature heat sources. Theoretical limit of such a thermoradiative (TR) cell was analyzed using the Shockley-Queisser framework established for photovoltaic devices.[16] This potential was demonstrated experimentally by Santhanam and Fan,[17] although the achieved efficiency was low.

To improve the efficiency of such devices, understanding the working principle of a thermoradiative cell is required. We first start by reviewing the operation of a conventional PV cell. When a PV cell is illuminated by a hot source, such as the sun, the p-n junction of the cell is driven out of equilibrium due to the radiative generation of excess electrons and holes. As shown in Fig. 1a, these excess electrons and holes flow towards the contacts on the n-type and p-type semiconductors, respectively. Under the open-circuit condition, the excess electrons raise the electron quasi-Fermi level in the n-type region and the excess holes lower the hole quasi-Fermi level in the p-type region. This leads to a quasi-Fermi level splitting, equivalently a positive open circuit voltage configuration in the p-n junction. When connected to a load, a negative current, which is opposite to an externally forward-biased diode, flows to the external load and leads to power generation. Hence, a PV cell works in the fourth quadrant in power generation mode while a forward-biased diode works in the first quadrant.

In contrast, a thermoradiative cell is heated up to a higher-than-environment temperature by a heat source. Thermally excited electrons in the n-type region and holes in the p-type region diffuse towards the space charge region and recombine radiatively. If the radiative recombination rate is faster than the radiative generation rate, more photons will be emitted from the cell than it receives from the ambient, resulting in a negative open-circuit voltage condition where a quasi-Fermi level difference is established in the p-n junction (Fig. 1b). When a load is connected externally, electrons flow into the n-type region and holes flow into the p-type region. In this power generation mode, the TR cell works in the second quadrant, while an externally reversely-biased diode works in the third quadrant (Fig.1c).

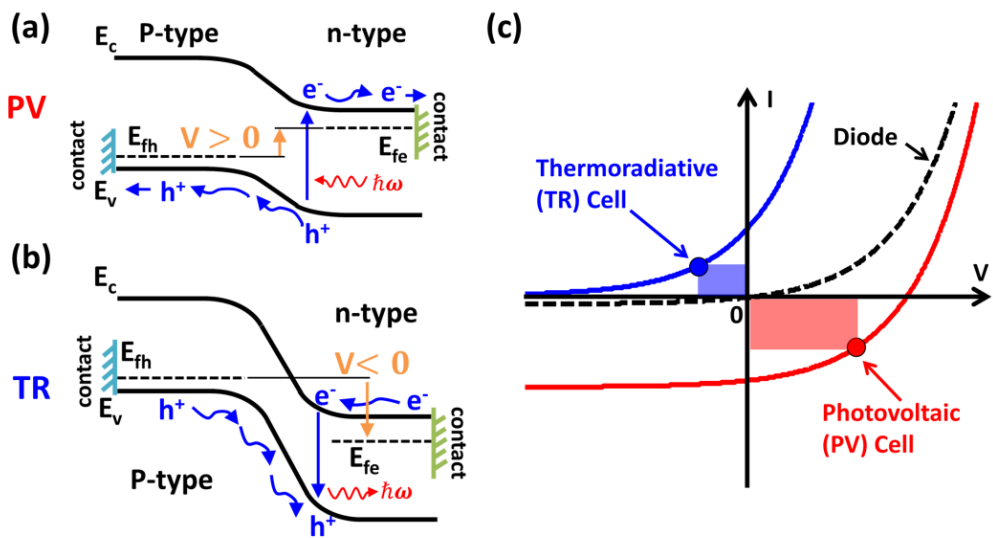

Fig. 1: (a) The band diagrams and (b) current-voltage curves of photovoltaic (PV) cells and thermoradiative (TR) cells. For a PV cell, excess electrons and holes generated radiatively by the sunlight will flow toward the contacts on the n-type and p-type semiconductors, respectively, resulting in a negative current in the load for power generation. On the contrary, in a TR cell, the thermally excited electrons and holes originating from non-radiative processes flow oppositely so that a positive current will flow to load and generate electricity.

Thermoradiative cell, as a heat engine, also needs to reject entropy carried in from the heat source as well as entropy generated in the device due to irreversible processes. To reject entropy to the environment, a thermoradiative cell relies on the out-going photons from thermal emission. The idea of using photons to reject heat of a heat engine is recently explored by Byrnes et al.[18] aiming at using outer space as the heat sink, and a thermoradiative cell can be a candidate for this approach.

Since thermoradiative cells rely on photons to carry out entropy, an understanding of photon entropy flux is important. In this paper, we start with an entropy analysis of photon flux. Guided by the analysis, we evaluate the performance of the TR cell with a spectrally-selective surface to demonstrate the device efficiency enhancement when the cell emissive spectrum is chosen to be narrow and centered at low frequencies. However, the increased device efficiency is accompanied by the reduction of its power density. This is a common dilemma for any energy converter, from the ideal Carnot engine[19] to photovoltaics,[9] thermoelectrics,[20] or thermophotovoltaics.[9,10-11] To overcome this dilemma, we propose engineering radiative energy transfer between the thermally emissive engine and the heat sink via coupling the generated electron-hole pairs to phonon polariton in the near-field.[21-22] The near-field coupling enables both narrow-band emission and a high enhancement of radiative energy transfer. The enhancement factor can be orders of magnitude higher than that of the far-field thermal emission.[21-23]

**Results**
**Entropy Analysis of Photon Radiation.** A heat engine using radiation to reject entropy such as a thermoradiative cell[16-18] has its efficiency bounded by the Carnot limit 1-$T_c$/$T_h$ due to the entropy generation caused by the imbalance of entropy flux between the heat conduction input, thermal emission, and light absorption from the environment. The origin of this imbalance is in the different amounts of entropy flux per unit energy in the processes of heat conduction (S/E=$T^{-1}$) and thermal radiation (S/E=4/3$T^{-1}$, for the blackbody emitter).[24-26] The study of the entropy of photons have a long history[24-27] and is well summarized in the book of Green.[9] To reveal the fundamental differences in the entropy content between radiation and heat conduction, the frequency dependence of entropy flux per unit energy is plotted in Fig. 2b by using the following expressions for the spectral energy and entropy fluxes of radiation, respectively:[25-26,28]

$$E(\omega) = \int_0^{2\pi} \int_0^{\frac{\pi}{2}} \frac{c}{4\pi} \varepsilon(\omega,\theta,\phi) \sin\theta \cos\theta D(\omega) f(\omega,T,\mu) \hbar\omega d\theta d\phi \quad (1)$$

$$S(\omega) = \int_0^{2\pi} \int_0^{\frac{\pi}{2}} \frac{c}{4\pi} \varepsilon(\omega,\theta,\phi) \sin\theta \cos\theta\, k_B D(\omega)[(1+f)\ln(1+f) - f\ln(f)] d\theta d\phi \quad (2)$$

Here, $E(\omega)$ is the photon energy flux per unit angular frequency, $S(\omega)$ is the entropy flux per unit angular frequency, $\varepsilon$ is the cooler surface emittance, $D(\omega) = \frac{n^3 \omega^2}{\pi^2 c^3}$ is the photon density of states in a homogeneous bulk material with refractive index n, $c$ is the speed of light, $f = [exp\left(\frac{\hbar\omega - \mu}{k_B T}\right) - 1]^{-1}$ is the photon distribution function, $\hbar$ is the Planck constant, $k_B$ is the Boltzmann constant, $n$ is the ambient refractive index, $\omega$ is the frequency of photons, $\theta$ and $\phi$ are the polar angle and azimuthal angle in the spherical coordinates respectively, and $\mu = E_{fe} - E_{fh}$ is the chemical potential of emitted photons. It should be noted that the chemical potential can be established when electrons and holes reach quasi-equilibrium among themselves to form quasi-Fermi levels for the electrons ($E_{fe}$) and holes ($E_{fh}$).[25,27] The entropy flux per unit energy is normalized to the heat conduction limit (S/E=$T^{-1}$), which is plotted as the black dotted line. The total (integrated over the whole frequency spectrum and angular range) normalized entropy content of the blackbody radiation equals 4/3, and is shown as the black dash-dot line.

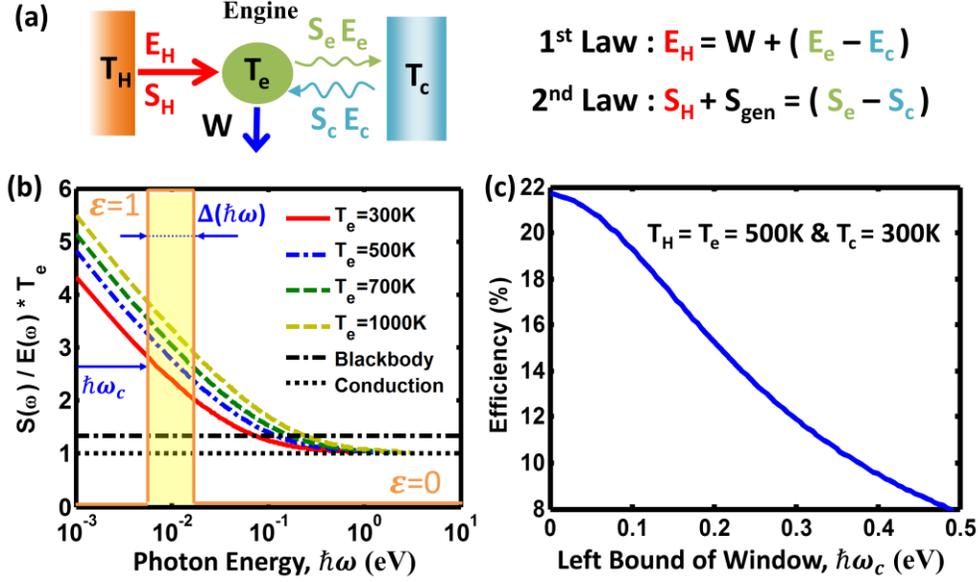

Figure 2: The thermally emissive engine. (a) The thermally emissive engine shows that the engine harvests energy ($E_H$) and entropy ($S_H$) via heat conduction from the hot side at temperature of $T_H$ to generate power (W), and the entropy is rejected through thermal emission ($E_e$ and $S_e$) from the engine at temperature $T_e$. The cold environment is at temperature of $T_c$. (b) The ratio of the spectral entropy flux to the energy flux reveals the entropy content per unit energy at various emissive temperatures and different photon energies. A band-pass selective surface with a width of the transparency window of $\Delta\hbar\omega$ is implanted to select the emissive photons. (c) Implanting a band-pass selective surface with a width of the transparency window of $\Delta\hbar\omega = 0.01$ eV, the efficiency increases when the low-frequency photons are selected for emission under the best scenario where no entropy is generated in the engine ($S_{gen}=0$). $\hbar\omega_c$ is the left bound of the band pass window. The input energy supplied in the form of heat conduction has entropy content of $S_H=E_H/T_H$.

Figure 2b provides a clue to increasing the converter efficiency via judicious choice of the channels to deliver the energy to the engine and to dump the excessive entropy to the environment. Heat conduction or high frequency radiation carry the lowest entropy content per unit energy, and thus are the most favorable forms of energy for the engine input. As work (W) carries zero entropy, the excessive entropy must be rejected to the heat sink. As illustrated in Fig. 2b, the best form of energy to maximize the entropy removal from the engine is low-frequency emission, which is characterized by the high entropy flux per unit energy. To verify this observation, we calculated the upper limit or the second law efficiency ($\eta_{2nd}$) for an engine, as shown in Fig. 2a, using heat conduction as the energy input channel (with the entropy content $S_H/E_H = T_H^{-1}$) and the thermal emission with a frequency-selective surface as entropy-dumping channel (with the entropy content $(S_e - S_c)/(E_e - E_c)$) under the best scenario where no entropy is generated in the engine ($S_{gen}=0$).

$$\eta_{2nd} = \frac{W}{E_H} = 1 - \frac{(E_e - E_c)}{E_H} = 1 - \frac{\frac{1}{T_H}}{\frac{S_e - S_c}{E_e - E_c}} \qquad (3)$$

It is known that entropy can be generated during emission and absorption processes.[9] This entropy generation, however, is not intrinsic and depends on the specific process. For example, the entropy generation can be reduced when the emitted photons acquire a chemical potential and vanishes when an ideal infinite junction solar cell works at open circuit condition.[9] As shown in Fig. 2b, the selective surface only allows emission within a narrow spectral window with a bandwidth of $\Delta\hbar\omega$ and completely blocks emission of photons outside this window (i.e., $\varepsilon(\hbar\omega) = 1$ within the window and $\varepsilon(\hbar\omega) = 0$ elsewhere, as schematically illustrated in Fig. 2b). The engine efficiency is plotted with varying lower bounds ($\hbar\omega_c$) of the band pass window. Given $\Delta\hbar\omega = 0.01$ eV and $S_{gen}$=0, figure 2c clearly shows that the second law efficiency can be enhanced when photon emission is selected to radiate out at low frequency.

**Principle of the thermoradiative cell.** As shown Fig. 1, a thermoradiative cell is thermally driven out of equilibrium, generating a reverse-bias condition due to emitting more photons than received from the environment so that the electrons and holes are collected by contacts on the p-type and n-type semiconductors, respectively. When an external load is connected to the TR cell, the thermally excited electrons and holes flow from n-type and p-type regions, respectively, and contribute to the current ($I$). The power (W= $-\mu I$) can then be generated by the TR cell at a given temperature and calculated from the equation of current ($I$) continuity using the following expression:[16]

$$I(\mu) = q[N(Te,\mu) - N(Tc,0)] \qquad (4)$$

where

$$N(T,\mu) = \int_{E_g}^{\infty} \varepsilon(\omega)D(\omega)cf(\omega,T,\mu)d\omega \qquad (5)$$

where chemical potential of photons ($\mu = E_{fe} - E_{fh} < 0$).[27] The TR cell efficiency ($\eta$) can then be calculated as $\eta$ = W / $E_h$, where $E_h = \int_{E_g}^{\infty} E(\omega)d\omega$ and $E(\omega)$ is integrated using Eq. (1).

We used Eqs. (1) - (5) to calculate the characteristics of a TR cell with a band gap ($E_g$) of 0.1eV, the cell temperature $T_H$=500K, and the ambient temperature $T_C$=300K. The efficiency and the power density (Fig. 3a) are shown for various chemical potentials of the TR cell, under the assumption that the emission only occurs as a result of the interband radiative recombination process. The TR cell emission spectra at different temperatures and voltages for this situation are plotted in Fig. 3b. The resulting device efficiency (Fig. 3a) has been previously defined by Strandberg[16] as the TR cell limit. By comparing the curves in Fig. 3a, we can observe that the maximum efficiency point (MEP) does not correspond to the same value of the chemical potential as the maximum power density point (MPP) due to the reason that the energy input of heat conduction also adjusts under different voltages as shown in Fig. 3c. In particular, the maximum efficiency is achieved when the power density is decreasing. Furthermore, the maximum TR cell efficiency is far below the Carnot efficiency limit 1-$T_c$/$T_h$.

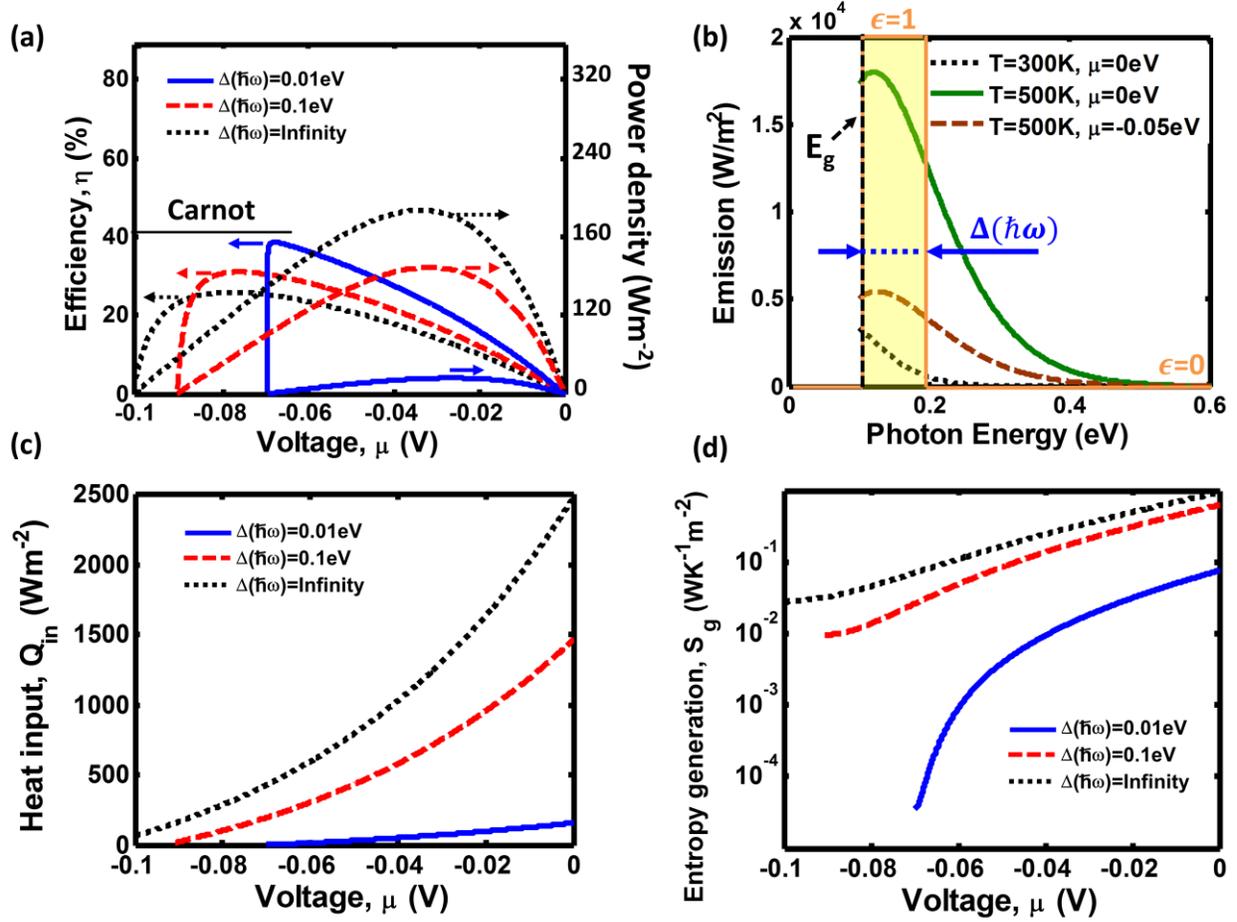

Figure 3: The performance of a thermoradiative cell with a band gap of 0.1 eV operating between a heat source at 500 K and a cold environment at 300 K. (a) Efficiencies and power densities of the TR cell are calculated as a function of various negative voltage and for three selected widths of the emission spectral window: $\Delta\hbar\omega= \infty$, $\Delta\hbar\omega= 0.1$ eV, and $\Delta\hbar\omega= 0.01$ eV. (b) Schematics of the narrowband spectral windows of the TR cell are shown overlapping with the emission spectra, and the emissive spectra are also calculated for varying cell temperatures and voltages. (c) Heat conduction input and (d) entropy generation are also calculated under the same operating conditions.

As we predicted based on the entropy analysis illustrated in Fig. 2a, the TR cell efficiency can be enhanced in the situation when its emission is highly spectrally selected to be narrowband and dominated by low frequency photons. Figure 3a also shows the calculated results when the emission window of a TR cell are limited to $\Delta\hbar\omega= 0.1$ eV and $\Delta\hbar\omega= 0.01$ eV above the band gap. We used these spectral windows to calculate characteristics of the TR cell with the spectrally-selective emission. Our data show that when low frequency photons dominate the emission spectrum, the maximum efficiency is increased and close to the 1-$T_c$/$T_h$ Carnot limit (Fig. 3a). However, the maximum power density generated by the cell with the narrowband frequency-selective emittance decreases as the spectral window width is reduced (Fig. 3a), resulting in the lower entropy generation as shown in Fig. 3d.

Based on the comparison between the efficiency and power density curves in Fig. 3a, we can better quantify the TR cell performance by replotting these two values together, as shown in Fig. 4. The power-efficiency curves begin at origin under zero chemical potential, corresponding to short circuit condition. When a load is connected, a negative voltage is biased on the load. As the magnitude of the voltage increases, the power density first increases and approaches the maximum power density point. Upon passing this point, the power starts to decrease but the efficiency keeps increasing until it reaches the maximum efficiency value. Finally, the curve folds and returns back to the origin due to a decreasing current. For the widths of the emission spectral window of $\Delta\hbar\omega = \infty$, $\Delta\hbar\omega = 0.1$ eV, and $\Delta\hbar\omega = 0.01$ eV, the power-efficiency curves are shown in Fig. 4 as black dashed line, red narrow solid line, and blue solid line, respectively. The maximum efficiency can be clearly observed to be increasing with the shrinking spectral width of the emission window, but the power density is reduced.

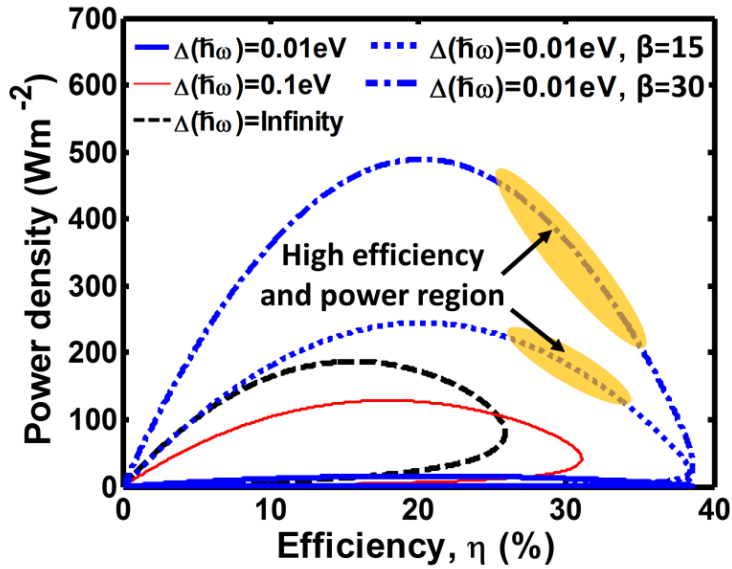

Figure 4: The power-efficiency curve of a thermoradiative cell with the band gap of 0.1 eV operating between a heat source at 500 K and a cold environment at 300 K. Power-efficiency curves reveal the trends of operating efficiencies and power densities. Blue dotted line and blue dash-dot line illustrate hypothetical power-efficiency loop if the emission to cold sink can be enhanced by a factor β. Such enhancements are possible for near-field TR cells.

The drop of power density stems from the reduced amount of emitted photons through a narrowband radiative channel. This imposes a limitation on the performance of the TR cell, when its thermal emission operates in the far field. However, it has already been demonstrated both theoretically[10,29-30] and experimentally[21-22] that the radiative power flux can be increased within a narrow spectral range by orders of magnitude if the emitter and the heat sink are electromagnetically coupled at a small gap distance through the near-field. To roughly estimate the effect of such enhancement, we introduced an enhancement factor (β) to emulate the near-field radiation within a narrow spectral window of $\Delta\hbar\omega$ (i.e., $E = \beta\varepsilon(\omega)D(\omega)\frac{c}{4\pi}f(\omega,T,\mu)\hbar\omega\Delta\hbar\omega$; $\varepsilon(\omega) = 0$ outside the emission window). The expressions for the first and second law in Fig. 2a can then be modified as:

$$E_H = W + (E_e - E_c)\beta \qquad (7)$$
$$S_H + S_{gen} = (S_e - S_c)\beta \qquad (8)$$

The power-efficiency curves for a TR cell with enhancement factors $\beta = 15$ and $\beta = 30$, restricted by the emission spectral window width $\Delta\hbar\omega = 0.01$ eV and operating at temperature $T_e$=500K, are shown as blue dotted line and blue dash-dot lines in Fig. 4. Comparison of the curve for the same cell without near-field enhancement ($\beta = 1$, solid blue line in Fig. 4) reveals opportunities to operate TR cells at high efficiency and high power density simultaneously.

**Near-field thin-film thermoradiative cell.** To further test our hypothesis, we built a detailed numerical model of a near-field enhanced TR cell. The near-field enhanced TR system consists of a thin-film TR cell and a semi-infinite radiation extractor (heat sink), which exchange radiative energy through a narrow vacuum gap. The choice of a thin-film TR cell stems from minimizing the electronic loss through the non-radiative processes, which will be discussed later, and the optical constants including InSb and $CaCO_3$ and thickness-dependent results are shown in the supplementary material. In order to utilize the enhanced and narrow-band near-field radiative energy transfer between the TR cell and the heat sink, the gap distance (g) should be on the order of or smaller than the dominant emission wavelength.[10,21-22,30] The submicron gap distance can enable efficient coupling of photons generated by the radiative recombination process in the TR cell with the phonon polariton on the surface of the radiation extractor (i.e., heat sink). Typically, phonon polariton modes have resonant frequencies below 0.2 eV, e.g., 0.186 eV (~ 6.7 μm) for $CaCO_3$ and 0.138 eV (~ 9 μm) for $SiO_2$.[21,30] On the other hand, InSb with a band gap of 0.17 eV at room temperature[32] is a semiconductor material commonly used to make narrow-gap PV cells and infrared detectors.[33-34] The phonon polariton mode of $CaCO_3$ has resonant frequency slightly higher than the band gap of InSb. It allows the phonon polariton coupling to electrons through the interband transition to direct extract the luminescent emission, making $CaCO_3$ material a good candidate for the narrow-band near-field radiative extractor. Other materials are also possible to be used as near-field radiative extractors, including materials supporting surface plasmons polaritons.[35]

To compare the performance of TR cells under various scenarios of radiative energy extraction, we calculate the power-efficiency curves for several systems, including: (i) the InSb thin film cell (Fig. 5a), (ii) the InSb thin film with a spectrally-selective surface (Fig. 5b), and (iii) the InSb thin film with the adjacent $CaCO_3$ acting as the near-field radiative extractor (Fig. 5c). First, we calculate an ultimate limit of the device efficiency and power density assuming that *the radiative process through the interband transitions is the only mechanism of radiative energy extraction* (ideal case). Such an assumption corresponds to the celebrated Shockley-Queisser efficiency limit for photovoltaic cells.[36] In Fig. 5d, the limiting power-efficiency curves are compared for the three systems. The energy carried away from the cell by photons with frequencies below the band gap energy (*sub-bandgap loss*) and the non-radiative generation of free carriers via Auger, Shockley-Read-Hall, and surface defect processes (*non-radiative loss*) are the two major loss mechanisms in a realistic system, and their impacts on the device performance will be discussed later.

First, to calculate the emission spectrum of a 50-nm-thick InSb thin film (Fig. 5a), which is close to the optimal thickness for each of the three configurations (see the thickness dependent results in the supplementary material), we used the transfer matrix method[37] to rigorously solve Maxwell equations to find the surface emittance, $\varepsilon(\omega)$. The TR cell is assumed to have a perfect metal mirror as a backside reflector. The emittance was then averaged over the total angular range from 0 to 90º, and over the two orthogonal (transverse electric and magnetic) polarizations of emitted photons using Eq. (1) and (2). Finally, using Eq. (4) and (5), the power and efficiency were calculated and plotted together as a black solid curve in Fig. 5d. The emission spectrum of the InSb thin film with a spectrally-selective surface is plotted in Fig. 5b. It can be seen by comparing Figs. 5a and 5b that not only the spectrum bandwidth but also the peak emission power flux significantly decreases for the TR cell with the selective surface. As a result, the device current density and power drop, but the efficiency increases as expected, shifting the power-efficiency curve to the region of low power density and high efficiency (red solid line in Fig. 5d).

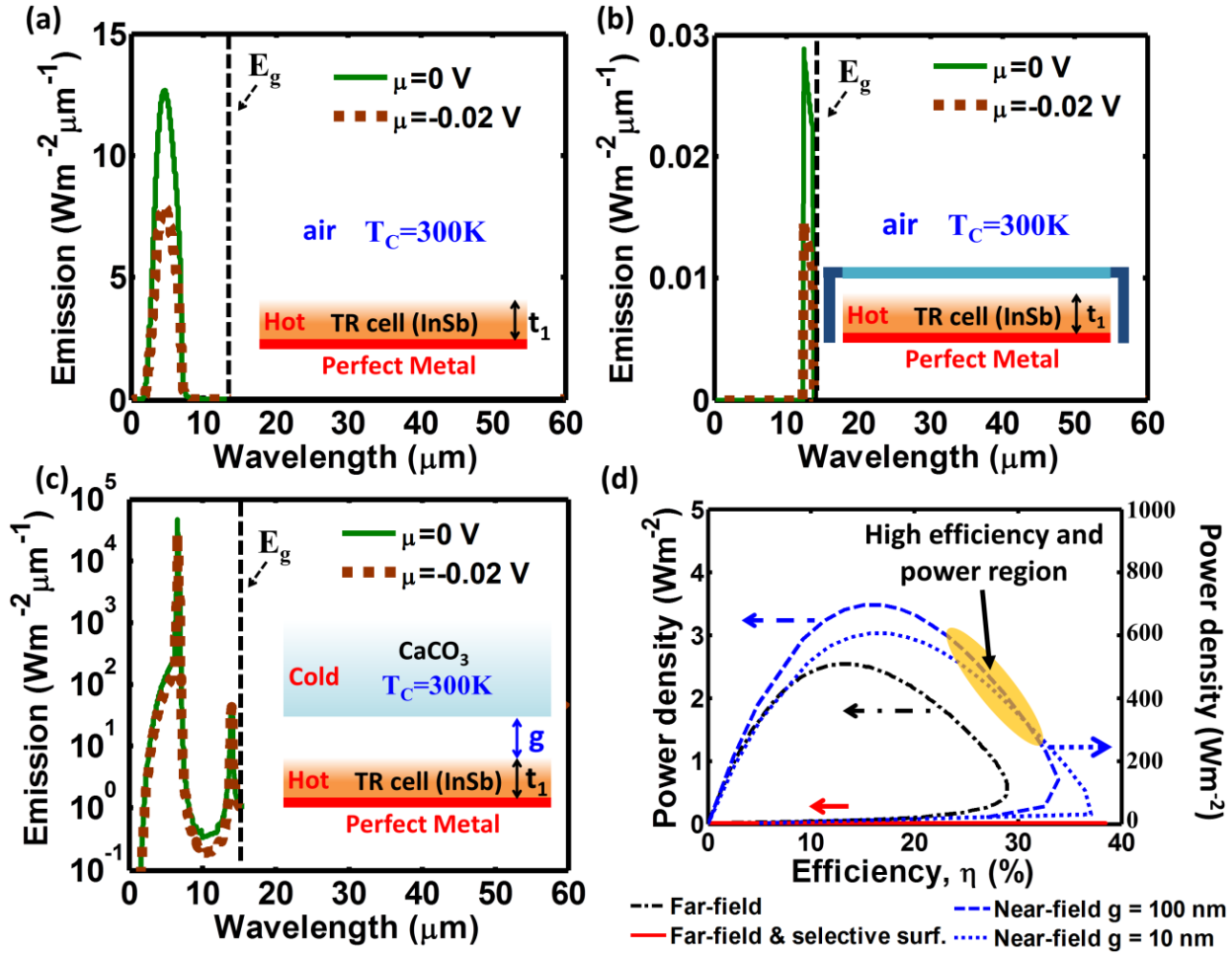

Figure 5: The performance of the 50-nm-thick InSb TR cells operating between a hot source at 500 K and a cold environment at 300 K, calculated for various scenarios of the radiative energy extraction. The emission spectra for the three scenarios, including (a) a far-field emission from a thin-film InSb cell, (b) a far-field narrowband ($\Delta\hbar\omega = 0.01$ eV) emission from a thin-film InSb cell with a selective surface (surf.), and (c) a near-field phonon-polariton-enhanced energy transfer from a thin-film InSb to a bulk $CaCO_3$ across 10nm and 100nm-wide vacuum gapS (g). (d) Power-efficiency curves are calculated for the three energy extraction scenarios, with the black dash-dot line, the red solid line, and the blue dotted and dashed lines representing the systems with the emission spectra shown in (a), (b), and (c), respectively.

The spectrum of the near-field radiative energy transfer from the InSb thin film to a bulk $CaCO_3$ is narrow-band and over two orders of magnitude enhanced in intensity over its far-field counterpart. Importantly, most of the photons contributing to the radiative heat flux have energy greater than the InSb bandgap energy. As such, they can significantly enhance $N(T_e, \mu)$ in Eq. (4), yielding high current and high power density of the TR cell. To calculate the near-field energy transfer from a thin film to a bulk medium via a narrow gap, a rigorous analytical electromagnetic formulation based on Rytov theory is used.[38] The thermally emitting object is modeled as a volume of fluctuating dipole sources, whose amplitudes are determined via the fluctuation dissipation theorem.[10,30,38-41] The detailed formulas can be found in the supplementary material and ref. 10 and 35. To generate the plots in Fig. 5c, the radiative energy transfer from a 50-nm-thick InSb thin film to a bulk $CaCO_3$ was calculated for a gap distance (g) of 10 nm. A narrow-band and enhanced radiative energy transfer can be observed, peaking at the phonon-polariton resonance wavelength of 6.7 µm as shown in Fig. 5c. As a result, the TR cell power-efficiency curve extends to the regions of both high power and high efficiency, which offers opportunities to efficiently harvest low-grade heat into electricity.

Although the near-field emission of the TR cell, using the coupling of the phonon polariton mode in the heat sink, can provide the narrow-band and intensity-enhanced features, it may also introduce additional losses into the system, such as sub-bandgap emission losses. The sub-bandgap emission can also strongly couple to phonon-polariton modes in the heat sink, leading to its heating and thus to degrading the efficiency of the TR system. For the case of the near-field radiative energy transfer between the InSb cell and $CaCO_3$ extractor, low-energy phonon polariton modes with wavelengths of 28 µm in $CaCO_3$ can contribute to the sub-bandgap losses. These losses are manifested as additional low-frequency peaks in the near-field heat flux spectrum in Fig. 6a.

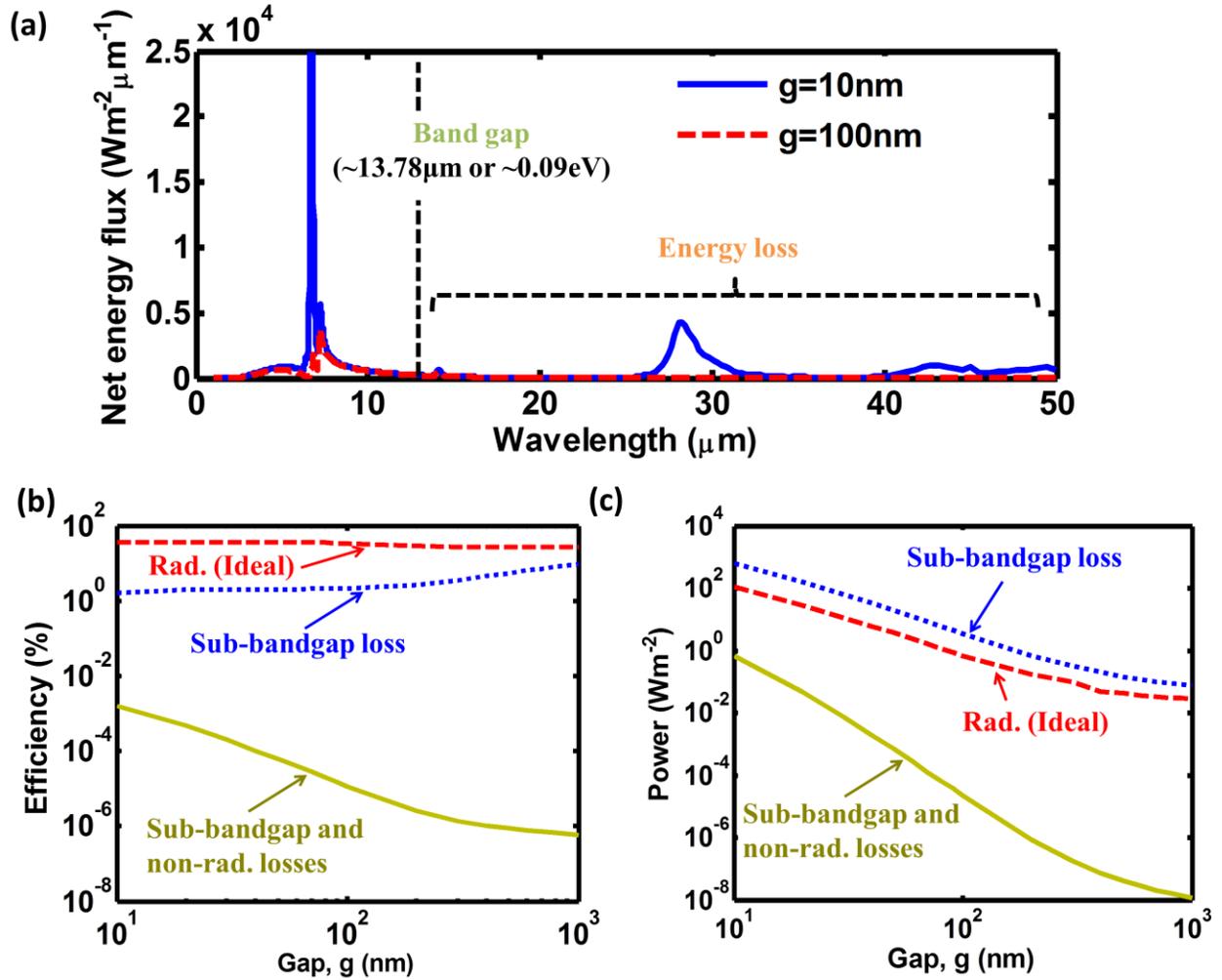

Figure 6: The effect of the sub-bandgap emission and non-radiative recombination losses on the near-field TR cell performance. The device is comprised of a 50-nm-thick InSb cell at $T_H$ and a bulk $CaCO_3$ extractor at $T_C = 300$ K. (a) Net near-field energy flux between InSb and $CaCO_3$. InSb bandgap narrows at elevated temperatures, and equals 0.09 eV at 500 K.[32] The device (b) efficiency and (c) power density are calculated at varying gap distances, assuming cell operation at maximum efficiency point. The blue plots for the ideal cases are shown for comparison. The brown plots include the sub-bandgap loss. The yellow plots consider both sub-bandgap and non-radiative losses (due to the Auger and surface defect processes).

The data in Fig. 6a show that there is a trade-off between increasing the above-bandgap emission and reducing the sub-bandgap losses. The best coupling configuration can be found by optimizing the thickness of the TR cell and the width of the vacuum gap between the cell and the extractor. We included the sub-bandgap loss mechanism into our near-field heat transfer model and calculated the system performance as a function of the vacuum gap distance (g) between the TR cell and heat extractor (Fig. 6b and 6c). We observe that the maximum efficiencies are reduced from the ideal cases when the sub-bandgap losses are taken into account (compare blue plots to brown plots in Fig. 6b). However, the corresponding power densities increase as shown in Fig. 6c due to the mismatch of the maximum efficiency point and the maximum power point.

**Discussion**
**Effect of Non-Radiative Losses.** Furthermore, realistic models of TR cells should include non-radiative losses. The non-radiative processes (Auger, Shockley-Read-Hall, and surface defects processes) contribute to the net free carrier generation in the TR cell and provide an additional loss mechanism in a realistic system. To account for the losses due to the Auger and Shockley-Read-Hall (SRH) processes, their rates need to be included in current continuity equation:[42-43]

$$\nabla I = q[G_{rad} - R_{rad}] + q[G_{Auger} - R_{Auger}] + q[G_{SRH} - R_{SRH}] \quad (9)$$

where $q$ is the elementary charge, the $G_{rad}$, $G_{Auger}$, and $G_{SRH}$ are the free carrier generation rates for radiative, Auger, and Shockley-Read-Hall processes respectively, and the $R_{rad}$, $R_{Auger}$, and $R_{SRH}$ are the free carrier recombination rates for radiative, Auger, and Shockley-Read-Hall processes respectively. Typically, the Shockley-Read-Hall process is neglected in InSb due to its slow rate compared to the radiative and Auger processes.[44-46] To account for the surface defect recombination process, the current continuity equation needs to be solved together with the Poisson and drift-diffusion equations to calculate the net free carrier generation rate (see supplementary material). After incorporating both sub-bandgap and non-radiative losses in InSb into our model, we observe that both efficiencies and power densities drop by orders of magnitude relative to their counterparts for the system with only radiative and sub-bandgap losses (compare yellow plots to brown plots in Figs. 6b and 6c). However, both efficiency and power density increase with the narrowing gap distance (Figs. 6b and 6c).

To provide guidelines for performance improvement of realistic TR cells, we need to identify the process that dominates the net generation of free carriers and contributes to the major energy loss. In Fig. 7a, the net rate of the inter-band radiative process and the net generation rates of the Auger and surface defect processes are calculated for varying gap distances. All rates are increased for smaller gap distances, but the radiative recombination rate increases faster than non-radiative generation rate, resulting in the increase of power density and efficiency. Obviously, the Auger process is characterized by a much faster rate than the surface defect process, and thus it dominates the energy loss through the free carrier generation. For the InSb, the ratio ($m$) of the radiative rate to the non-radiative rate (including both Auger and surface defect rates), $m$ = |radiative rate| / |Auger rate + surface defect rate|, is calculated around 2~2.5 at each gap distance. If the suppression of Auger process can be achieved in either InSb or another material to improve this ratio to m > 4, the major loss through non-radiative generation can be significantly reduced. As a result, the device efficiency can be increased by more than three orders of magnitude (Fig. 7b). The ongoing research efforts to suppress the Auger process using quantum wells,[47] doping profile control,[48] superlattices,[49] or heterostructures,[50] can yield performance improvement of TR cells.

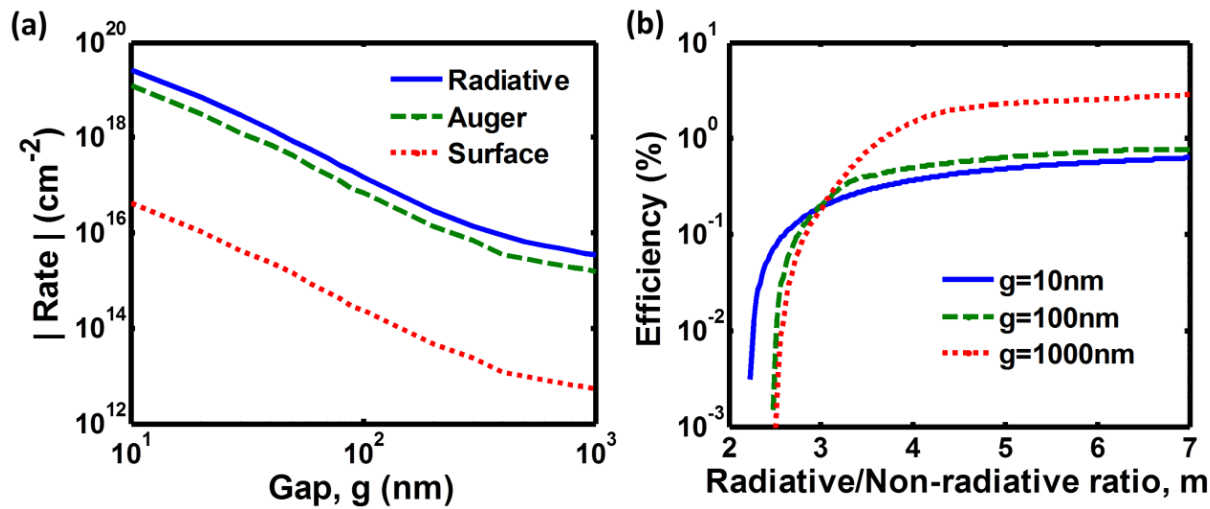

Figure 7: Non-radiative losses in a near-field TR cell comprised of a 50-nm-thick InSb film at $T_H$ = 500 K and a bulk $CaCO_3$ extractor at $T_C$ = 300 K. (a) Net rates of the free carrier generation and recombination due to radiative, Auger, and surface defect processes are calculated as a function of the gap distance. The ratio (m) of the radiative rate to non-radiative rate in InSb is calculated to be m≈ 2~2.5 in this configuration for all gap distances. (b) Device efficiency can be enhanced by five orders of magnitude provided that this ratio can be increased to m > 4.

**Conclusion**

To summarize, we predict that the efficiency of thermoradiative (TR) cells can be dramatically improved if they are engineered to selectively emit only low-frequency photons. This prediction is based on a theoretical analysis of the entropy content of high- and low-frequency radiation. Spectral selectivity of far-field emission can be achieved by introducing a selective surface, which boosts the efficiency of a TR cell but abate the power density due to the reduced emission and a lower radiative recombination rate. This work show that near-field radiative energy transfer between the TR cell and the heat sink, which is enhanced by resonant coupling between electron interband transition on the hot side and phonon polariton modes on the cold side, can simultaneously yield both high efficiency and high power density, offering opportunities to efficiently harvest low-grade heat. However, the sub-bandgap radiation losses and the non-radiative losses, especially due to the Auger process, may degrade the performance of TR cells significantly. For example, in a realistic system composed of a 50-nm-thick InSb TR cell at 500 K and a $CaCO_3$ heat sink at 300 K coupled via a 10nm vacuum gap, the performance of this TR system can be improved from $2.4 \times 10^{-5}$ % and $10^{-5}$ Wm$^{-2}$ (far-field) to 0.5 % and 45.16 Wm$^{-2}$ (near-field at 10 nm gap distance). Under this configuration, the ratio of the radiative rate to the non-radiative rate is calculated around 2, which causes to significant non-radiative losses. Provided the sub-bandgap radiation loss can be eliminated, the performance can further achieve 20.4 % and 327 Wm$^{-2}$ from 2.5 % and 161 Wm$^{-2}$, when the nonradiative-to-radiative ratio is improved from 2 to 15.3.

**Acknowledgement**
The authors gratefully acknowledge helpful discussions with Professor Nicholas X. Fang and Professor Jesus A. del Alamo, the Massachusetts Institute of Technology, USA. This work is support in part by DOE BES (DE-FG02-02ER45977) (near field radiation, W.-C. H., J.K. T., Y. H., and G. C.) and by DOE BES EFRC (S3TEC) under DE-SC0001299/DE-FG02-09ER46577 (for heat to electrical energy conversion, B. L., S.V. B. and G. C.)

**Additional information**

**Competing financial interests:** The authors declare no competing financial interests.

**Author contribution**
W.-C. H. came up with the idea to use low-frequency photons and near-field radiation to improve thermoradiative cells. W.-C. H., J. T., B. L., Y. H., S. B., and G. C. contribute to numerical calculations, results discussion, and paper writing. S. B. and G. C. supervised this project.